# Unveiling Voices: A Co-Hashtag Analysis of TikTok Discourse on the 2023 Israel-Palestine Crisis

## Digital Methods & Information Analytics

***Report Info:***

| | |
|---|---|
| **Course Unit:** | Research Report 1 |
| **Name:** | Rozin |
| **ID:** | Hasin |
| **Course Instructor(s):** | 12567582 |
| | Dr. Marissa Wilcox |
| **Research Report Title:** | Unveiling Voices: A Co - Hashtag Analysis of TikTok Discourse on the 2023 Israel-Palestine Crisis |
| **Word Count:** | 2193 |

**I acknowledge that I have followed the university's Plagiarism guidelines and have properly attributed parts of this work sourced from elsewhere.**

Yes

**I give permission for this work to be used as an example of good work in future editions of the course.**

Yes

***Report Contents:***

# 1. Introduction

On October 8th, the State of Israel declared war officially on Hamas for the first time since the 1973 Yom Kippur War (Cenciotti and D'Urso), in retaliation to the Hamas-led attack on the Supernova Music Festival a day earlier. While it is impossible to remain politically neutral, it is possible to look at concrete facts - almost 20,000 civilians murdered in 2 months (Sawafta and Fick). This cataclysmic crisis is being experienced by the world both offline, through sit-ins, walkouts, mobilizations and protests in major cities around the globe, and online, through the lenses of the people on the ground posted on social media platforms – particularly TikTok, which has secured its spot as the top for short-form content among the youth. TikTok, which since its release in September 2016 has effectively been China's answer to the social media hegemony of the West, serves as a lens through which to examine how users engage with, interpret, and contribute to discussions surrounding this catastrophic ongoing crisis. As a platform known for its byte-sized content, TikTok presents a distinctive challenge and opportunity to its users in regards to participation in online discourse, particularly surrounding contested topics such as geopolitical conflicts.

Last year, Abbas et al. published a study on how TikTok's affordances were used to share public opinion on a previous escalation of the conflict in Sheikh Jarrah in 2021. The study concluded that TikTok's affordances encouraged the creation of viral public content, making the platform a space for political expression, mobilization, and activism (Abbas et al.). Reflecting on this study and to reaffirm its main finding, I intend to study TikTok, through its recommendation algorithm rather than its affordances, and investigate its role in respect to the ongoing crisis.

In this study, I investigated how political content on the crisis is dealt with and disseminated by TikTok's recommendation algorithm - what kind of results it returns when people search for information about the conflict? What is being recommended concerning the crisis on TikTok? Is that a balanced view of the issue or is it biased or limited in some way? By implementing a co-hashtag analysis on 509 videos obtained from the results of the recommendation algorithm based on key search terms such as "#israel" and "#palestine", this study adds to the current literature on TikTok's pervasive algorithm and its role in stimulating online activism. In terms of social implications, this study aims to contribute to the ongoing debate on the crisis and its relation and manifestation in digital spaces.

# 2. Question

What kind of information does the Tik Tok recommendation algorithm show in its search results concerning the ongoing Israeli occupation of Gaza through querying the terms "#israel" and "#palestine"?

# 3. Method

I situate my research design in between discourse analysis, which is an approach to the analysis of written, vocal, or sign language use, or any significant semiotic event, and algorithm studies, which is the critical study of the social, cultural and political life of the algorithm and its conditions of change, evolution and possibility (Jenna Ng and David Theo Goldberg). I argue that discourse analysis can be extended to the form of mixed media content prevalent on TikTok. In my research, I am particularly interested in the notion of "search as research" (Taibi et al.) in respect to TikTok, as I will be looking into the metadata (hashtags) of content returned by searching a particular political term. According to the Lomborg and Kapsch notion of modes of engagement with algorithms (Lomborg and Kapsch), my research design is centered around a "preferred" mode of engagement with the TikTok algorithm.

The following practical steps form the core methodological approach for my research design:

1. **Software Installation**: To collect the data from TikTok, the first necessary step is to install the Zeeschuimer extension [provided by the Digital Methods Initiative ( https://tools.digitalmethods.net/zeeschuimer ) on a suitable research browser, configure it to capture TikTok posts and link the preferred 4CAT server where the harvested data will be sent to (look at later steps).
2. **Fresh Slate**: To make sure that any previous use of the platform does not produce any bias in the research process, it is necessary to either clear the account data of an existing account or make a fresh account on TikTok.
3. **Data Collection**: I queried the following - "#israel" and "#palestine" - simultaneously and collected the top 251 and 258 posts for the respective hashtags through Zeeschuimer.
4. **Processing**: The next step is to visualize the collected data through 4CAT. 4CAT provides a lot of "processors" to manipulate and interpret the data, which is also available in .ndjson and .csv formats.
5. **Network**: To draw a hashtag network, the Co-tag network operation was used on the dataset to obtain a .GEXF network file of tags co-occurring in a posts. Edges are weighted by the amount of tag co-occurrences; nodes are weighted by how often the tag appears in the dataset. To make the network more readable for our purposes of answering our question, the Circle Pack Layout was applied to obtain a readable network, adjusting some parameters such as edge visibility and node size, and coloring the nodes using the Louvain Community color scheme.

6. **.csv Data**: To further analyze the most important nodes and edges of the network, the .csv data is filtered and interpreted for the most viral content. Some of the filtering can be done directly in 4CAT itself through the given processors.

# 4. Analysis

The aim of my research was to uncover the particular kinds of information a user might fight on the occupation in Gaza by querying key terms into TikTok's search algorithm. What I found is that the content is more political than not, and that there is a variance of political leanings in the content. In this section, I have tried to summarize the key findings of my network analysis of this content.

The obtained co-hashtag network is made up of 9 communities (Table 1). The communities have been coded by color, and the table below represents the related categories of hashtags for each community, roughly from largest to smallest community.

| Community | Category | | |
|---|---|---|---|
| | pro-palestine | | |
| | pro-israel | | |
| | TikTok-specific | | |
| | sports and media | | |
| | pro-palestine | | |
| | pro-palestine with mentions of qatar | | |
| | anti-war | | |
| | pro-IDF | | |
| | mostly unrelated | | |

*Fig 1: Communities in the hashtag network*

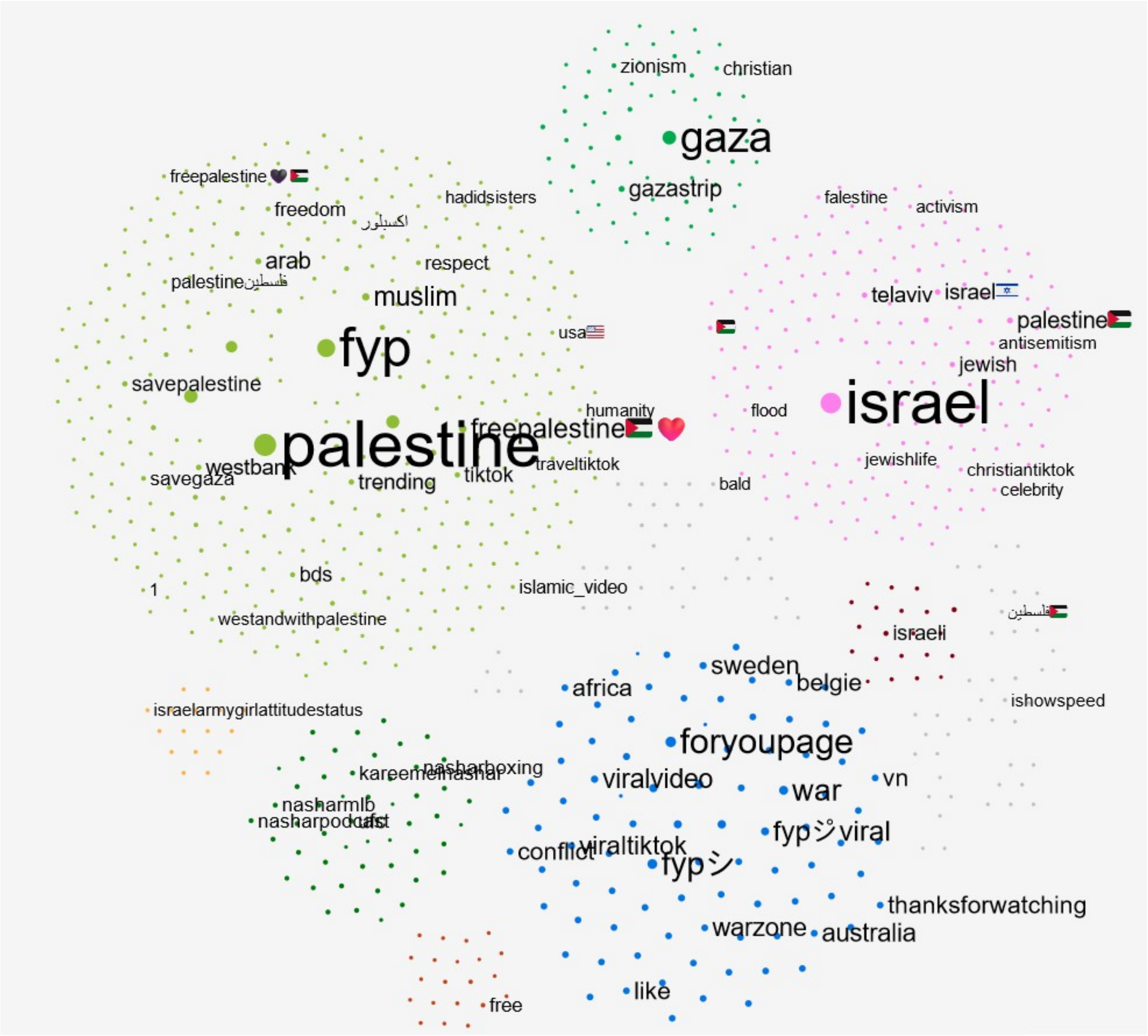

*Graph 1: Co-Hashtag Network labelled by node size*

This is the obtained hashtag network, as visualized in 4CAT using the Circle Pack layout. The largest and most dense community is made up of pro-Palestine hashtags, colored in light green. The pink and blue communities follow closely, consisting of pro-Israel hashtags and those that are TikTok-specific or general global terms (africa, sweden, belgie, australia, vn). The first thing to note is that "#palestine" is more connected than "#israel" (Table 3 and 4, Appendix). To look more closely at the network, I turn to the hashtag "#freepalestine❤" (Graph 2). It was found 37 times, in our dataset of 509 TikToks, and is very useful in seeing the connectedness of the message frames present in our dataset. The use of the heart symbol in conjunction with the slogan "free palestine" is meant to convey an emotional appeal to the audience, engaging them and possibly mobilizing them for online and offline action. Closely connected is another similar hashtag: "#freepalestine🖤", where a black heart can be argued to signify solidarity. This hashtag is also connected to other hashtags which are important in the geopolitical context: “#israel”, being the aggressors in this genocide, but also “#usa”- which is not surprising as the US government is biggest supporter of the genocide, ideologically, strategically and financially – and other pro-Palestine stances like “#savegaza”, “#savepalestine”, “#weststandswithpalestine”.

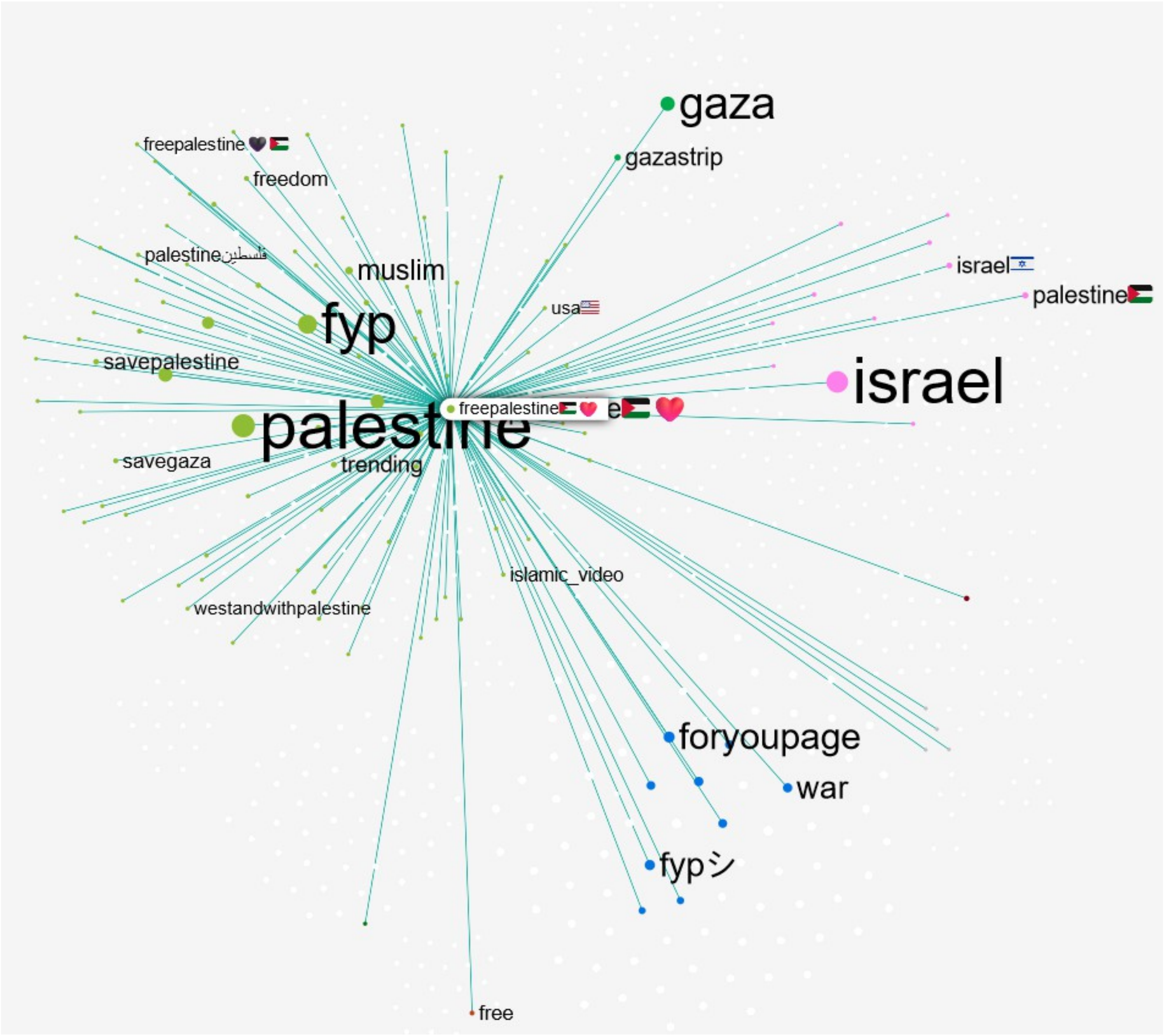

*Graph 2: Connections of #freepalestine*❤

Next, let's look at the top 50 hashtags that appeared in our dataset, given in Table 2. Firstly, it is important to note that the two most used hashtags were actually our queries - “#israel” (199) and “#palestine” (192). Next, we can see that there are some common hashtags at the top of the list that are TikTok-specific : “#fyp” (118), “#viral” (58), and “#foryou”(34). These hashtags are common practice by users who hope to move their content further up the recommendation algorithm and land on the For You page.

| date | item | value |
|---|---|---|
| all | israel | 199 |
| all | palestine | 192 |
| all | fyp | 118 |
| all | freepalestine | 102 |
| all | gaza | 89 |
| all | foryou | 58 |
| all | freepalestine🇵🇸❤️ | 37 |
| all | viral | 34 |
| all | news | 33 |
| all | foryoupage | 30 |
| all | war | 27 |
| all | muslim | 25 |
| all | hamas | 24 |
| all | fypシ | 22 |
| all | fy | 17 |
| all | palestine🇵🇸 | 17 |
| all | jerusalem | 15 |
| all | islam | 15 |
| all | arab | 14 |
| all | idf | 13 |
| all | military | 12 |
| all | soldier | 11 |
| all | israeli | 11 |
| all | فلسطين | 11 |
| all | telaviv | 11 |
| all | israel🇮🇱 | 11 |
| all | palestinian | 11 |
| all | trending | 10 |
| all | gazaunderattack | 10 |
| all | gazastrip | 10 |
| all | palestinetiktok | 10 |
| all | jewish | 9 |
| all | muslimtiktok | 9 |
| all | army | 9 |
| all | freepalestine🇵🇸 | 9 |
| all | fypage | 8 |
| all | palestina | 8 |
| all | savepalestine | 8 |
| all | protest | 7 |
| all | capcut | 7 |
| all | visitisrael | 7 |
| all | tiktok | 7 |
| all | standwithisrael | 6 |
| all | 🇮🇱 | 6 |
| all | cnn | 6 |
| all | fypシviral | 6 |
| all | israelwar | 6 |
| all | ufc | 6 |
| all | fromtherivertotheseapalastinewillbefree | 6 |
| all | palestinewillbefree | 6 |

*Table 2: Top 50 hashtags in the dataset*

Now, looking beyond these first two categories, we see a prevalence of pro-Palestine hashtags: "#freepalestine" (102), "#freepalestine❤" (37), "#gazaunderattack" (10) , "#palestine" (9), "#savepalestine" (8), "#fromtherivertotheseapalastinewillbefree" (6), and"#palestinewillbefree" (6) for instance. Pro-Israel hashtags are fewer, while being comparatively lower in frequency: "#hamas" (24), and "#standwithisrael" (6) for instance.

This reflects that overall, the pro-Palestine narrative is stronger, accompanied by users using the affordances of the TikTok to promote the narrative, compared to support for Israel, indicating the instrumentality of grassroot voices in this crisis.

# 5. Discussion

Moving beyond the quantitative findings, there are some important takeaways from the data qualitatively. The most liked TikTok, with 5.9m likes, is by an influencer GagGeniusHQ who has 730k followers, showing a montage of clips of destruction in Gaza after an IDF attack. This shows that influencers on the platform who were otherwise not political in their content used their huge fan bases to push political content at the time and raise awareness about horrific incidents happening. This is a novel phenomenon of the online activism paradigm (Abbas et al.), and shows the power and influence that micro-celebrities have over their audiences in their ability to generate awareness and spread information.

Another phenomenon is quite interesting. Ranked close to the top 10 most liked videos, the 12th most liked video is a handheld clip of a young female Israeli soldier walking by in military gear. The video appears like a handheld shot from a passerby. Accompanying the conventional pro-israel hashtags is #beautiful. This kind of video is part of a larger trend of videos that are part of a particular community in our dataset: TikToks of women in the IDF. These TikToks have been noted to have a particularly fetishizing quality to them, and a vast majority of them are posted by official IDF channels. While the IDF is not new to using social media as a propaganda tool (Dickson), it is certainly unique how they weaponize their own women to impact the narrative.

What is particularly significant is the temporality of the hashtags. Table 3 shows the frequency of the collected hashtags over time. The initial bump around April 2021 corresponds to the Sheikh Jarrah attack that was the focus of the study published by Abbas et al. After that, there is a sharp dip with almost no hashtags for almost a year, which I think reflects the time period when the attention on major social platforms was on the COVID-19 pandemic. Hashtags start appearing around mid 2022, and gradually increase in 2023. The sharpest rise was in October 2023, when the conflict escalated with the events of the 7th. What is interesting to note is that the frequency becomes 1/3rd of October in November, despite the real-world conflict getting worse, without a hint towards a ceasefire – revealing that social media news cycles are very short lived compared to the duration of real-world events: even though it appears that people are initially heavily invested, they move on pretty quickly to other topics and a form of desensitization occurs.

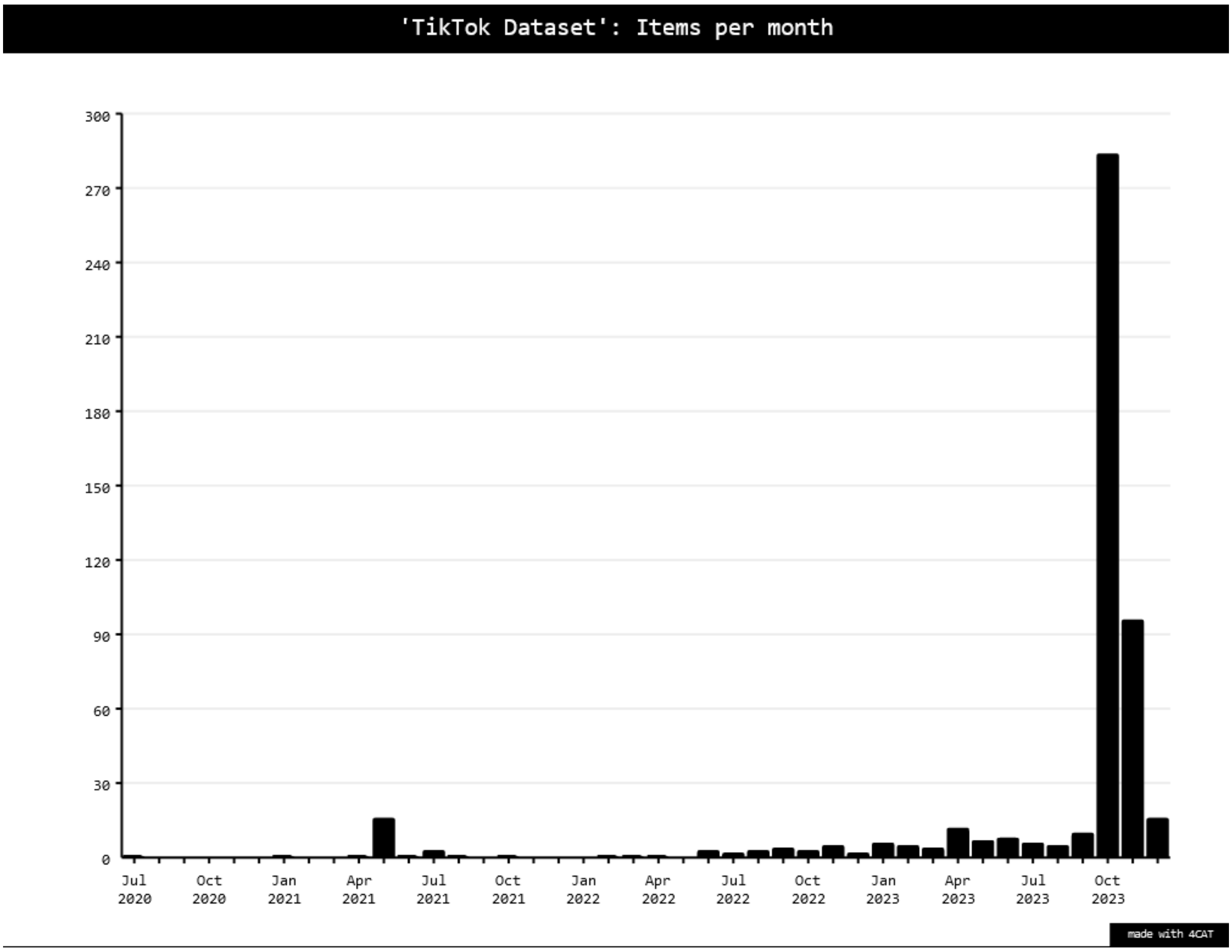

*Table 3: No.of hashtags per month*

This follows previous research, which has shown that it is difficult for social movements, particularly protests voicing concern regarding global conflicts, to gain legitimacy in the mainstream news cycle, where protests, such as the pro-Palestine ones we are currently seeing erupt in cities around the world, are covered as 'episodic', ignoring their larger issues (Poell).

# 6. Conclusion

In this research, I started out by questioning what kind of information TikTok shows us regarding the ongoing genocide when we search the platform using terms like #israel and #palestine, like anyone wanting to inform themselves on current events. Through my research I have shown how young people are using TikTok to spread a pro-Palestine narrative by harnessing the affordances of the medium. I have discussed the phenomenon of adapting political content in the social media activism paradigm and the weaponization of certain groups for propaganda. I have also argued towards the temporality of geopolitical crises in news cycles.

This research design is specific in context and uses a small sample. Future methodologies can look into working with bigger samples of data. Cross-platform analysis would also be beneficial for mapping the issue across a variety of platforms, thereby providing a platform-independent lens into the cumulative media being circulated. Fluency in Arabic would also add to the analytical depth, as more regional and local content could be contextualized. It would also be interesting to look into the comments of the top posts to study interactivity and calls for mobilization among audiences. Finally, looking into the background of prominent influencers promoting pro-Palestinian or pro-Israeli content could be helpful in gauging their positioning and stakes in the conflict.

It is crucial that this kind of research into the ongoing crisis continues, given the gravity of the situation which demands us as researchers to analyze and document this period in history for the future generation of academics to come, so that the general public remains rightfully informed of what is going on, which is one step towards collective action.

# 8. Reflection

This paper was firstly a great academic exercise, and secondly a meaningful personal journey. While I got to learn how to work with technical software to find answers to my research questions, I also needed to grasp the seriousness of the work in light of the current times. There was often disturbing content that I encountered, which is part of the research process. I greatly improved my analytical skills in working with data through this project. As I do not use TikTok myself, it was interesting to look at it from a research perspective and analyze the presence of a social issue on it. I also gained historical knowledge about the Israel-Palestine conflict through doing background research on the issue. The experience was meaningful and I hope this work inspires other researchers to carry out similar projects through which they can make their voices heard.

# 9. Appendix

Graph 3: Connections for #palestine

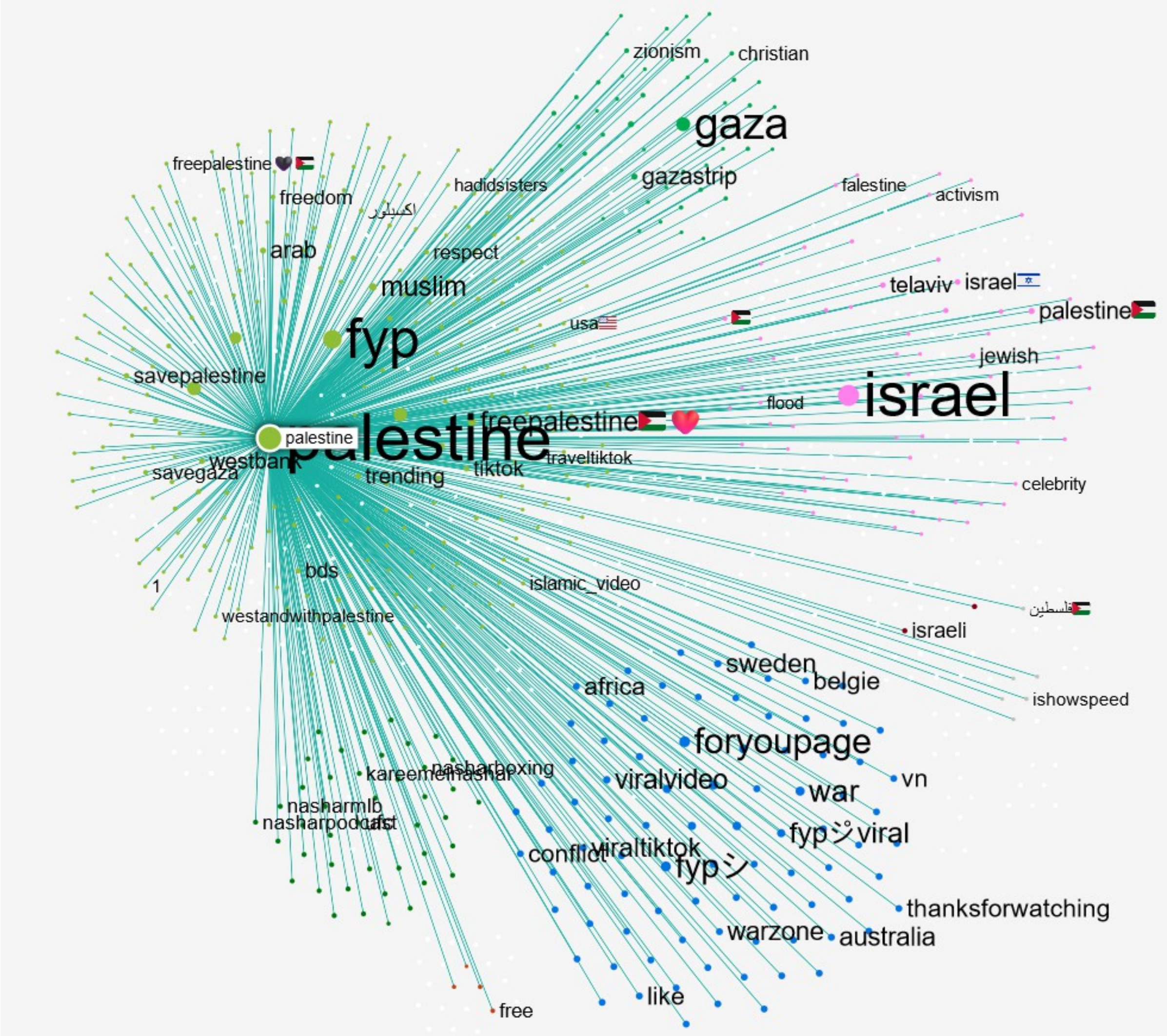

Table 4: Connections for #israel

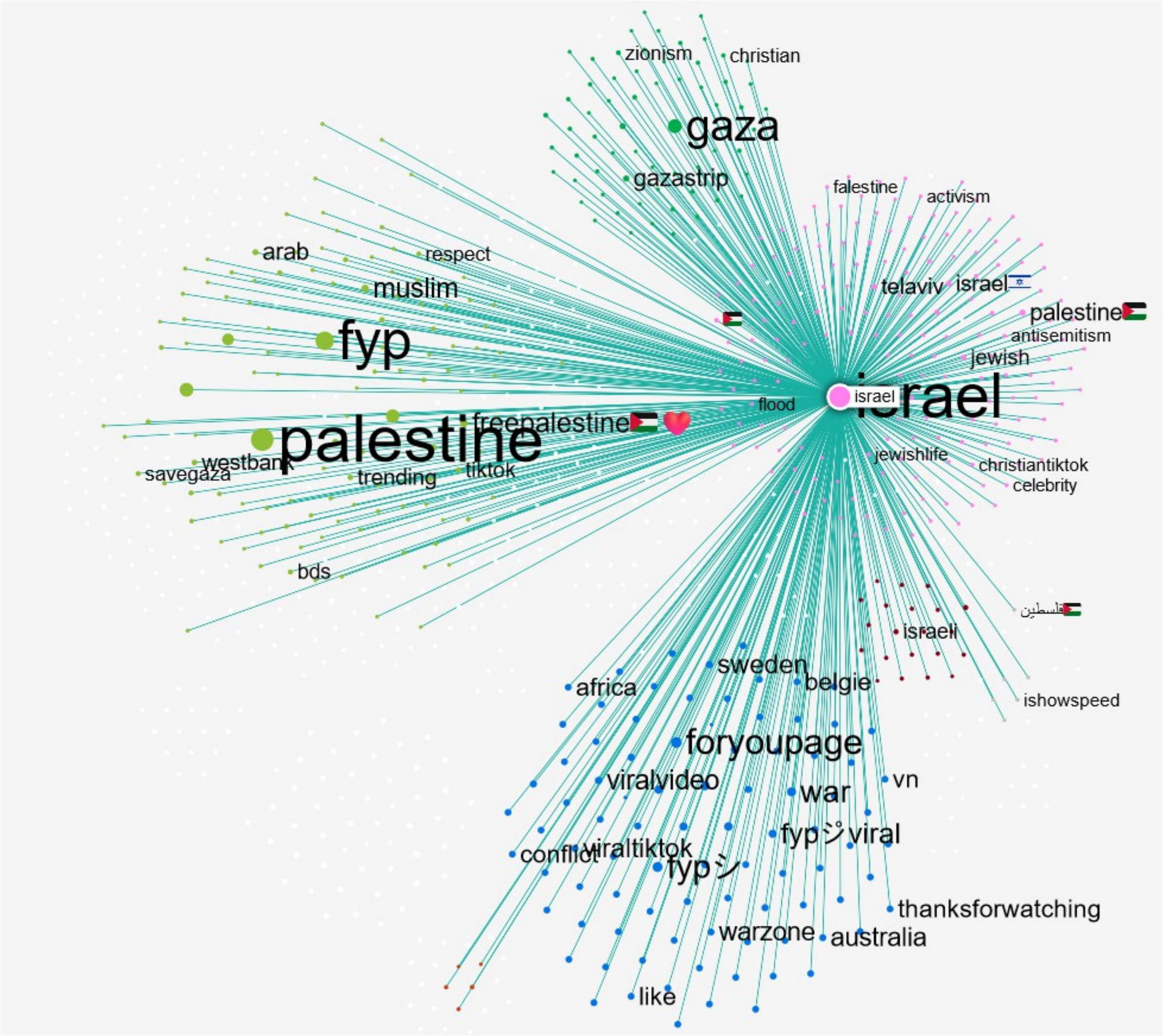